\documentclass[journal]{IEEEtran}
\usepackage{amsmath,amsfonts}
\usepackage{algorithmic}
\usepackage{algorithm}
\usepackage{array}
\usepackage[caption=false,font=normalsize,labelfont=sf,textfont=sf]{subfig}
\usepackage{textcomp}
\usepackage{stfloats}
\usepackage{url}
\usepackage{verbatim}
\usepackage{graphicx}
\usepackage{cite}
\usepackage{hyperref}
\usepackage{ulem}
\usepackage{soul}
 \hypersetup{
     colorlinks=true,
     linkcolor=black,
     filecolor=black,
     citecolor = black,      
     urlcolor=black,
     }

\usepackage{physics}
\newcommand\numberthis{\addtocounter{equation}{1}\tag{\theequation}}

\begin{document}

\title{Modeling the 3-micron Class Er-Doped Fluoride Fiber Laser with a Cubic Energy Transfer Rate Dependence}

\author{William Bisson, Alexandre Michaud, Pascal Paradis, Réal Vallée, and Martin Bernier}

\markboth{IEEE Journal of Quantum Electronics}%
{Shell \MakeLowercase{\textit{et al.}}: A Sample Article Using IEEEtran.cls for IEEE Journals}


\maketitle

\begin{abstract}
We propose an energy transfer model with a cubic atomic population dependence to accurately model the behavior of various reported high-power erbium-doped fluoride fiber lasers operating near 2.8~microns. We first show that the previously introduced weakly interacting (WI) and strongly interacting (SI) models are not adequate for precisely modeling such high-power erbium-doped fluoride fiber lasers. We compare results obtained with the WI and SI models to the proposed model by simulating 4 different highly doped (7~mol.\%) fiber lasers previously reported in the literature. Laser efficiencies and powers are reproduced with great accuracy. In addition,
four other  fiber laser systems based on erbium  concentrations varying from 1-6 mol.\% are also simulated with good accuracy using the proposed model with the exact same set of spectroscopic parameters, which  confirms its validity for various erbium doping concentrations. Redshifting of laser wavelength is also taken into account by considering the full cross section spectra and computing signal powers over several wavelength channels. 
\end{abstract}


© 2024 IEEE.  Personal use of this material is permitted.  Permission from IEEE must be obtained for all other uses, in any current or future media, including reprinting/republishing this material for advertising or promotional purposes, creating new collective works, for resale or redistribution to servers or lists, or reuse of any copyrighted component of this work in other works.\\
DOI: 10.1109/JQE.2024.3372580

\section{Introduction}
\IEEEPARstart{H}{igh}-power erbium-doped fluoride fiber lasers operating near 2.8~microns are undergoing rapid development due to their various high-end  applications, for instance in biomedicine, material processing and remote sensing \cite{reviewmir}. A commonly used pumping band for these lasers is the erbium transition from the fundamental level $^4I_{15/2}$ to level $^4I_{11/2}$, which is around 976~nm \cite{faucher, martin, ozan, fortin, tokita, licolasing, ozancolasing}. The Stokes efficiency of erbium lasers pumped in this band is around 33\%, but it was found that colasing at 1.6~$\mu m$ can help increasing this efficiency to 50\% \cite{ozancolasing}. The Stokes efficiency was also exceeded by using highly doped systems, which led to higher energy transfer upconversion rates, and an optical-to-optical efficiency of 35.4\% was reached by Faucher et al. \cite{faucher}.  More recently, the record CW power of 42~W, with an overall efficiency of 22.9\%, was reported by Aydin et al. \cite{ozan} by pumping from both ends of the laser cavity and by improving the fiber laser assembly. 

\begin{figure}[H]
\centering
\includegraphics[width=3.3in]{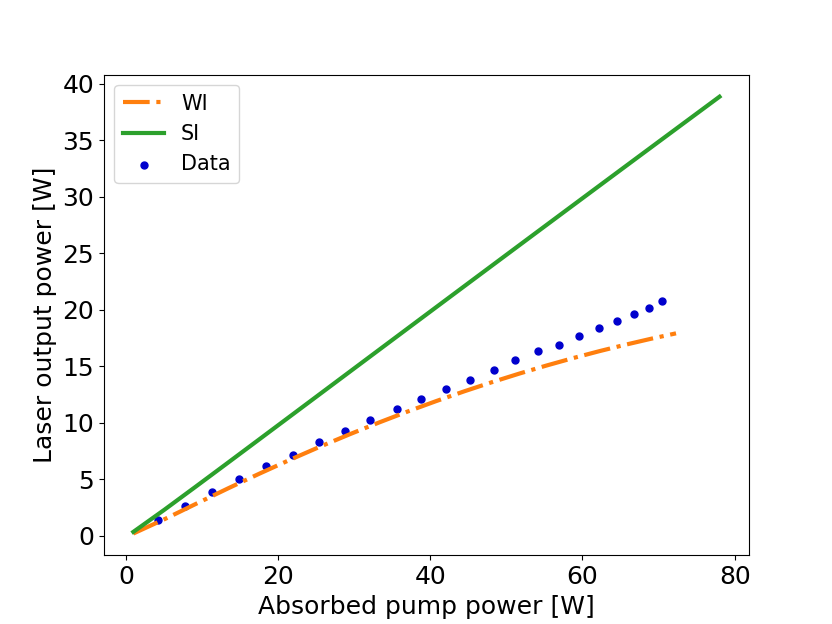}
\caption{Simulation of the 7 mol.\% 2825 nm laser from ref.~\cite{faucher} with the WI and SI models and  a pump wavelength of 976~nm.}
\label{fig:faucherwisi}
\end{figure}

The modeling of erbium-doped fluoride fiber lasers operating around 2.8~microns has been a subject of interest for multiple years \cite{pollnau, lisim, guo, gorjan}, but no satisfactory model was found to fit the multiple reported lasers with the same set of spectroscopic parameters. Having a precise model is important since it can guide  experiments and namely help scaling up output powers. It can also give access to underlying information, like atomic populations for example \cite{gorjanchap13}. A certain flaw actually exists in the literature regarding the energy transfer rates, for which two sets of parameters were proposed. These sets are named the weakly interacting (WI) and the strongly interacting (SI) models \cite{lisim}. The SI model relies on transfer rates that were derived from bulk samples whereas the WI model uses the same general equations but with scaled down values for the transfer rates. In addition, these sets of transfer rates change with the total erbium concentration \cite{pollnau}, which means they have to be measured every time a new concentration is used. The WI model was first introduced by Gorjan et al. \cite{gorjan} to obtain better agreement with experimental values, while Li and Jackson used the same set of energy transfer parameters to simulate multiple lasers \cite{lisim}. Although the WI model provides a better agreement with the experiment than does the SI model, it is however unable to reproduce the experimental lasing efficiency properly at all pumping powers. This can be seen in figure \ref{fig:faucherwisi} where we show simulation results of the 20~W laser from ref.~\cite{faucher} using the WI and SI models. The simulations in ref.~\cite{lisim} were also unable to accurately reproduce  the efficiency of this laser at all pump powers.
Both the SI and WI models rely on the same quadratic dependence of the ETU processes as originally proposed by Grant \cite{grant}. It assumes infinitely fast energy migration \cite{agazzi} and has been shown to be inappropriate in many experimental cases \cite{zubenko}. 

In this paper, we propose an alternative handling of the energy transfer mechanism for the accurate modeling of high-power diode pumped erbium-doped fluoride fiber lasers operating around 2.8~microns. This approach is based on previously reported models \cite{dexter, bartolo, agazzi, jacksonchap6} and combines the extreme cases of energy transfer without the assistance of migration and energy transfer that is purely assisted by migration. We show that the proposed energy transfer model, which has a cubic dependence on excited atomic level population, provides better agreement with experimental data than the WI and SI models for 4 different heavily-doped (7~mol.\%) fiber lasers reported in the literature \cite{faucher, ozan, martin, fortin}. We show that such model can also be used, with the exact same set of spectroscopic parameters, for modeling a 6~mol.\% laser  \cite{tokita}, an unpublished 5~mol.\% laser and two colasing (at 1.6 and 2.8~microns) systems based on 1.5  and 1~mol.\% erbium-doped fluoride fibers respectively \cite{licolasing, ozancolasing}. The proposed model reproduces  the efficiency drop in accordance to pump power which is generally observed in heavily erbium-doped fluoride fiber lasers with good accuracy, and is thus suitable for predictions related to power scaling of such laser systems. It does not consider the clustering of ions, which still has not been directly proven for erbium-doped fluoride fiber lasers \cite{cluster}. We believe the new insight provided by this alternative modeling could pave the way to the accurate design of fiber laser systems operating over the 100W-level at 2.8 microns.

\section{Theoretical Model}

\subsection{1.6 microns and 2.8 microns erbium model}

\begin{figure}[!t]
\centering
\includegraphics[width=3.6in]{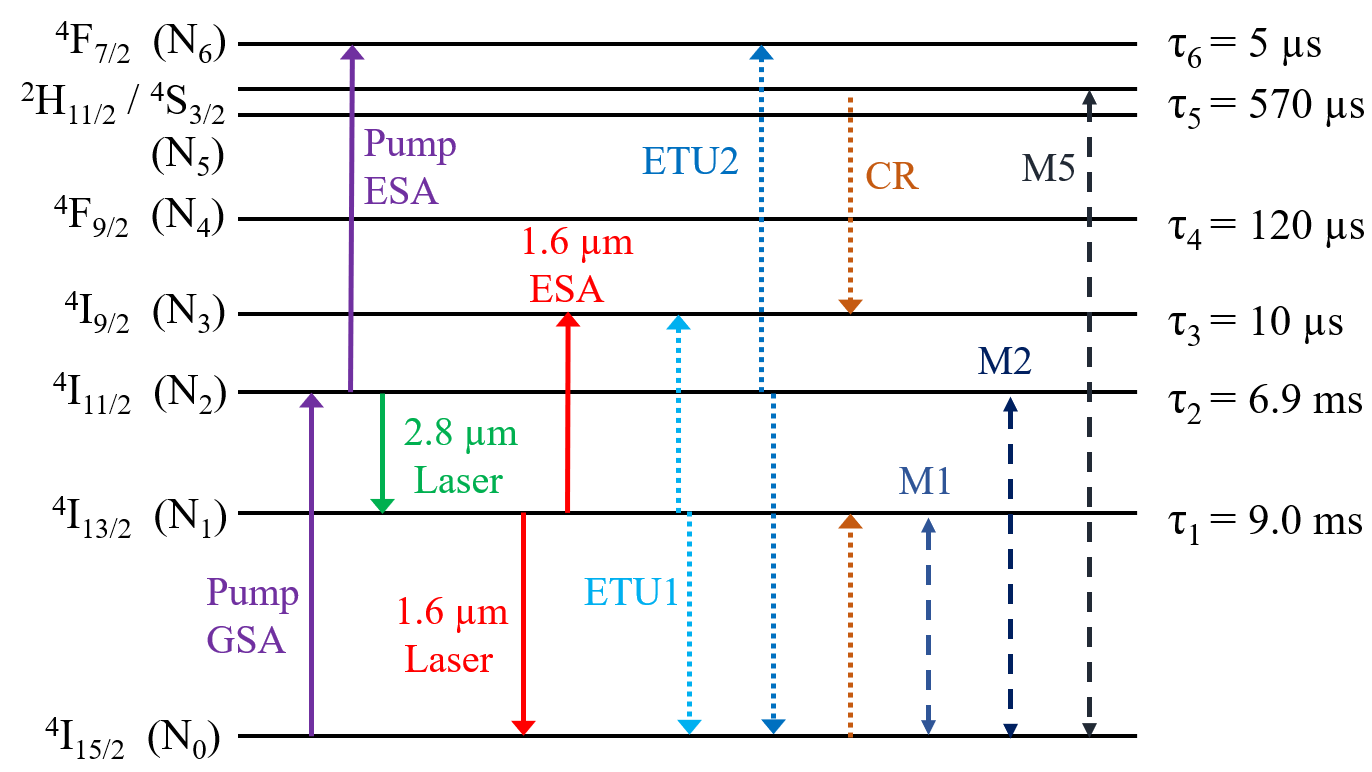}
\caption{Energy level diagram and lifetimes of erbium-doped fluoride glass. GSA = Ground State Absorption,  ESA = Excited State Absorption, ETU = Energy Transfer Upconversion, CR = Cross Relaxation, M = Migration. }
\label{fig:erpop}
\end{figure}

The energy level diagram containing the relevant radiative transitions and energy transfer processes for the proposed model is shown in figure \ref{fig:erpop}.

The atomic populations are calculated using the following set of rate equations \cite{lisim, guo}:
\begin{align*}
    \dv{N_6}{t} &= - N_6\tau_6^{-1} + R_{esa1}  + f_{etu2}, \numberthis \label{eq:niv6}\\
    \dv{N_5}{t} &= \beta_{65} N_6 \tau_6^{-1} - N_5\tau_5^{-1}  - f_{cr}, \numberthis  \label{eq:niv5}\\
    \dv{N_4}{t} &= \sum_{j=5}^6 \beta_{j, 4} N_j \tau_j^{-1} - N_4 \tau_4^{-1}, \numberthis  \label{eq:niv4} \\
     \dv{N_3}{t} &= \sum_{j=4}^6 \beta_{j, 3} N_j \tau_j^{-1} - N_3 \tau_3^{-1} + R_{esa2} \\ &+ f_{cr} + f_{etu1},  \numberthis  \label{eq:niv3} \\
     \dv{N_2}{t} &= \sum_{j=3}^6 \beta_{j, 2} N_j \tau_j^{-1} - N_2 \tau_2^{-1} + R_{gsa}  - R_{esa1} \\ &- R_{se28} - 2 f_{etu2}, \numberthis  \label{eq:niv2} \\
     \dv{N_1}{t} &= \sum_{j=2}^6 \beta_{j, 1} N_j \tau_j^{-1} - N_1 \tau_1^{-1}  - R_{se16}  + R_{se28} \\ &- R_{esa2} + f_{cr} - 2 f_{etu1}, \numberthis  \label{eq:niv1} \\
     N_{tot} &= \sum_{j = 0}^6 N_i, \numberthis  \label{eq:ntot} 
\end{align*}
which is solved in the steady-state regime by setting $\dv{N_i}{t} = 0$.  The atomic populations are then time independent $(N_i(z))$. 
$\tau_i$ are the lifetimes and  $\beta_{ij}$ are the branching ratios. 
Their values  for level 1 to level 5 were taken from \cite{lisim}, and values for $\tau_6$ and $\beta_{6j}$ were taken from \cite{guo}. 

The pump transition rates around  976~nm  are given by: \begin{align}
    R_{gsa} &= \sum_{\lambda} \xi \qty[N_0 \sigma_{02} - N_2 \sigma_{20}] [P_+ + P_-] \Delta\lambda,  \label{eq:gsa980} 
    \\
    R_{esa1} &= \sum_{\lambda} \xi \qty[N_2 \sigma_{26} - N_6 \sigma_{62}] [P_+ + P_-]\Delta\lambda. \label{eq:esa980} 
\end{align}
The laser signal transition rates around 2.8 $\mu$m and 1.6 $\mu$m are given by:
\begin{align}
    R_{se16} &= \sum_{\lambda} \xi \qty[N_1 \sigma_{10} - N_0 \sigma_{01}] [P_+ + P_-]\Delta\lambda, \label{eq:se1600} 
     \\
    R_{esa2} &= \sum_{\lambda} \xi \qty[N_1 \sigma_{13} - N_3 \sigma_{31}] [P_+ + P_-]\Delta\lambda, \label{eq:esa1600} 
    \\
    R_{se28} &= \sum_{\lambda} \xi \qty[N_2 \sigma_{21} - N_1 \sigma_{12}] [P_+ + P_-]\Delta\lambda, \label{eq:se2800} 
\end{align}
Where $se$ stands for stimulated emission. The sum is over all spectral channels, $\Delta\lambda$ is the spectral length of each channel and $P_+$ and $P_-$ are the power spectral densities in W/m  propagating in the forward and backward directions in the fiber respectively.  $P_+$ and $P_-$ are separated in an array containing a power density value for each wavelength channel. $\sigma_{ij}$ are the wavelength dependent absorption and emission cross sections. The $\xi$ factor is also wavelength dependent and is given by:
\begin{equation}
    \xi = \frac{\lambda \Gamma}{h c A_{\text{core}}},
\end{equation}
   where $\lambda$ is the wavelength of the considered channel, $A_{\text{core}}$ is the fiber core area and $\Gamma$ is the power-filling factor. 
For pump signals, which propagate in the cladding, the power-filling factor is given by the ratio of the core area to the cladding area of the fiber:
\begin{equation}
    \Gamma = \frac{A_{core}}{A_{clad}} = \frac{r_{core}^2}{r_{clad}^2}.
\end{equation}
For signal propagating in the core of the fiber, the power filling factor is:
\begin{equation}
    \Gamma = 1 - \frac{u^2}{V^2}\qty(1 -  \frac{K_{\nu}^2(w)}{K_{\nu+1}(w)K_{\nu-1}(w)}),
\end{equation}
where $u$, $w$ are the core and cladding transverse parameters, respectively, $V$ is the normalized frequency of the fiber and $K_{\nu}$ is the modified Bessel function of the second kind of order $\nu$.
In our simulations of colasing systems, two wavelength ranges, spanning from 1.4~$\mu$m to 1.7~$\mu$m and from 2.6~$\mu$m to 3.0~$\mu$m, were used. For lasers that only emit around $2.8~\mu$m, only the range from  2.6~$\mu$m to 3.0~$\mu$m was used, which means the optical transitions around 1.6~$\mu$m were not considered for these lasers. These power channels were separated in increments of $\Delta\lambda = 0.5$~nm. Using many channels provides the ability to let the system converge by itself to a lasing wavelength, instead of setting it as an input parameter \cite{gorjanchap13}. 

The pump absorption cross sections $\sigma_{02}$ and $\sigma_{26}$ were taken from \cite{csn0n2n2n6} and their respective emission cross sections ($\sigma_{20}$ and $\sigma_{62}$) were calculated using McCumber theory \cite{mccumber, rangemccumber} by using the energy values from \cite{starklevelserbium}. For the laser signal cross sections, $\sigma_{10}$ and $\sigma_{01}$ were taken from \cite{csn0n1n1n0}, $\sigma_{13}$ and $\sigma_{31}$ from \cite{guo}, while $\sigma_{21}$ and $\sigma_{12}$ were taken from \cite{csn1n2n2n1gsa}. The full cross section spectra were used for the simulations. 

According to the currently accepted energy transfer model, the three transfer rates are given by:
\begin{align}
    f_{etu1}^{G} &= W_{11} N_1^2, \label{eq:etu1grant} \\
    f_{etu2}^{G} &= W_{22} N_2^2, \label{eq:etu2grant}  \\
    f_{cr}^{G} &= W_{05} N_0 N_5, \label{eq:crgrant} 
\end{align}
Where $W_{11}$, $W_{22}$ and $W_{05}$ are constants  given in ref. \cite{lisim}. The $N^2$ dependence was derived by Grant \cite{grant} considering infinitely fast energy migration \cite{agazzi}.

The power propagation equation is given by:
\begin{equation}
    \dv{P_{\pm}}{z} = (g - \alpha) P_{\pm}
\end{equation}
where $P_{\pm}(z)$ represent the power spectral densities in W/m propagating in forward or backward directions, $g$ and $\alpha$ are the gain and  background loss at position $z$ and a given wavelength channel respectively and $dz$ is the length of each fiber section, which was set to 4~cm or less depending on the simulated system.  The gain is given by:
\begin{equation}
    g = \Gamma \sum_{\text{rad}} (\sigma_{ji} N_{j} - \sigma_{ij} N_i),
\end{equation}
where the sum is on all radiative transitions. Spontaneous emission was considered by adding a single photon in each wavelength channel \cite{se}.

\subsection{Energy Transfer Model}

According to Dexter \cite{dexter}, the energy transfer rate between two ions is given by the following equation \cite{agazzi}:
\begin{equation}
    W_{\text{DD/DA}} = \frac{C_{ik, j\ell}}{d^6}.
    \label{eq:transfertrate}
\end{equation}
Equation \eqref{eq:transfertrate} assumes electric dipole-dipole  interactions, which is the dominant process in the case of mid-infrared photonics  \cite{jacksonchap6, bartolo}. $d$ is the dipole distance and $C_{ik,j\ell}$ the Förster microparameter given by \cite{agazzi}:
\begin{equation}
    C_{ik,j\ell} = \frac{3c}{8 \pi^4 n^2} I_{ik,j\ell}.
    \label{eq:microp}
\end{equation}
$I_{ik,j\ell}$ is the overlap integral of the process and is given by:
\begin{equation}
    I_{ik,j\ell} =  \int \sigma^e_{ij}(\lambda) \sigma^a_{k\ell}(\lambda) d\lambda,
    \label{eq:overlap}
\end{equation}
where $\sigma^e_{ij}$ is the cross section of the emitting ion going from level $i$ to $j$ and $\sigma^a_{k\ell}$ is the cross section of the absorbing ion going
from level $k$ to $\ell$.
Equation \eqref{eq:transfertrate} to \eqref{eq:overlap} are valid for any energy transfer process, and overlap integrals for ETU1, ETU2, CR and migration processes are shown in table \ref{table:overlap}. The value for the CR process was guessed by comparing energy transfer values with the WI model since the $\sigma_{53}$ cross section is unknown.
\begin{table}[!t]
\caption{Overlap integrals   associated with each transfer process used in the theoretical model and references for the  cross sections used. }
\centering
\begin{tabular}{c c c c}
\hline
Process & $I_{ik,j\ell}$ &  Overlap [10$^{-58}$m$^5$] & Ref\\
\hline
ETU1  & $I_{10, 13}$ & 1.0 & \cite{guo, csn0n1n1n0}  \\
ETU2 & $I_{20, 26}$ & 3.1 & \cite{csn0n2n2n6} \\
M1 & $I_{10, 01}$ & 101 & \cite{csn0n1n1n0} \\
M2  & $I_{20, 02}$ & 4.5 & \cite{csn0n2n2n6} \\
CR  & $I_{53, 01}$ & 0.7 & \text{Guessed}  \\
M5  & $I_{50, 05}$ & 2.8 & \cite{csn1n2n2n1gsa} \\
\hline
\label{table:overlap}
\end{tabular}
\end{table}

According to Zubenko \cite{zubenko}, Grant's energy transfer model \cite{grant} does not apply in many experimental cases, since it assumes an infinitely fast energy migration process. As can be seen in figure \ref{fig:faucherwisi}, we were unable to reproduce the efficiency drop which was observed in ref. \cite{faucher} with a quadratic dependence of the energy transfer rate. We believe this points out the fact that Grant's model does not apply to the  3-microns class erbium-doped fluoride fiber laser system. Other excited level dependencies were thus tested by simulating several high-power  fluoride fiber lasers found in the literature \cite{faucher, ozan, martin, fortin, tokita, licolasing, ozancolasing}. Simulation results for all of these lasers are shown in the next section. 
 It was found empirically (cf. Appendix for more detail) that the following energy transfer rates for ETU1, ETU2 and CR gave the most accurate simulation results:
\begin{align*}
    f_{etu1} &=  a^{-6} C_{etu1}  N_1^3 \\ &+ \sqrt{\frac{\pi}{6}} a^{-6}  \sqrt{C_{etu1} C_{m1}}  N_0^{1/4} N_1^{11/4}, \numberthis \label{eq:fulletu1}\\
    f_{etu2} &=  a^{-6} C_{etu2}  N_2^3 \\ &+ \sqrt{\frac{\pi}{6}} a^{-6}  \sqrt{C_{etu2} C_{m2}}  N_0^{1/4} N_2^{11/4}, \numberthis \label{eq:fulletu2}\\
    f_{cr} &=  a^{-6} C_{cr}  (N_0 N_5)^{3/2} \\ &+  \sqrt{\frac{\pi}{6}} a^{-6}  \sqrt{C_{cr} C_{m5}}  (N_0 N_5)^{3/2}, \numberthis \label{eq:fullcr}
\end{align*}
where $C_{etu}$, $C_{cr}$ and $C_{m}$ are the Förster microparameters for ETU, CR and migration processes respectively and $a~=~\qty(\frac{4 \pi}{3})^{-1/3}$. It was found that adding a correction factor of 3\% to $a$ gave the best results. Equation \eqref{eq:fulletu1} to \eqref{eq:fullcr} can be applied in the concentration range of at least 1~mol.\% to 7~mol.\%, which will be shown in the next section. The cubic dependence of the energy transfer rate was found by assuming an uniform distribution of ions \cite{wangsimkin}. The first term was inferred directly from Dexter's theory \cite{dexter} and considers the donor to acceptor transfer process without the assistance of migration \cite{jacksonchap6}. The second term was inferred by using the hopping regime of purely migration assisted energy transfer and was calculated from the transfer rate given in ref. \cite{bartolo}. Combining these two terms to obtain the total transfer rates was found to be necessary in order to obtain accurate simulation results. 

The first terms of equations \eqref{eq:fulletu1} to \eqref{eq:fullcr} do not consider migration and are always nonzero as long as some ions are excited. The second terms were added to consider the cases when the migration process is also involved. For those terms, either the hopping regime or the diffusion regime applies. For the hopping regime to apply, the condition $\frac{C_{etu, cr}}{C_{m}} << 1$  must be verified \cite{zubenko}.  For the diffusion regime to apply, the  condition $\frac{C_{etu, cr}}{C_{m}} >> 1$  must be verified.
 Using equation \eqref{eq:microp}, we get:
\begin{equation}
    \frac{C_{etu, cr}}{C_{m}} = \frac{I_{ik,j\ell}^{\text{etu,cr}}}{I_{i'k',j'\ell '}^{\text{m}}}
\end{equation}
Using the overlap integral values in table \ref{table:overlap}, we get a ratio of 0.01 for ETU1, 0.7 for ETU2 and 0.25 for CR. Although the ratio is close to one for ETU2 and CR, the hopping regime was still applied.  If the diffusion regime was applied to these transfer rates, the impact on simulation results would be minimal. In addition, the hopping regime is very common in rare earth doped systems \cite{zubenko}.    For the second terms of the ETU transfer rates, the $N_0$ dependence comes from the donor concentration dependence of energy migration transfer rate \cite{agazzi, zubenko} and was found to be necessary in order to obtain accurate simulation results. Details on the derivation of equations \eqref{eq:fulletu1} to \eqref{eq:fullcr} can be found in the appendix.

In the next section, we show that the proposed energy transfer model provides improved modeling results as compared to the currently accepted WI and SI models, especially at high pump powers in heavily-doped fibers, by simulating 8 different laser systems, all with the exact same set of spectroscopic parameters.

\section{Results and Discussion}

\subsection{7  mol.\% erbium-doped fiber lasers}

\begin{figure}[!t]
\centering
\includegraphics[width=3.3in]{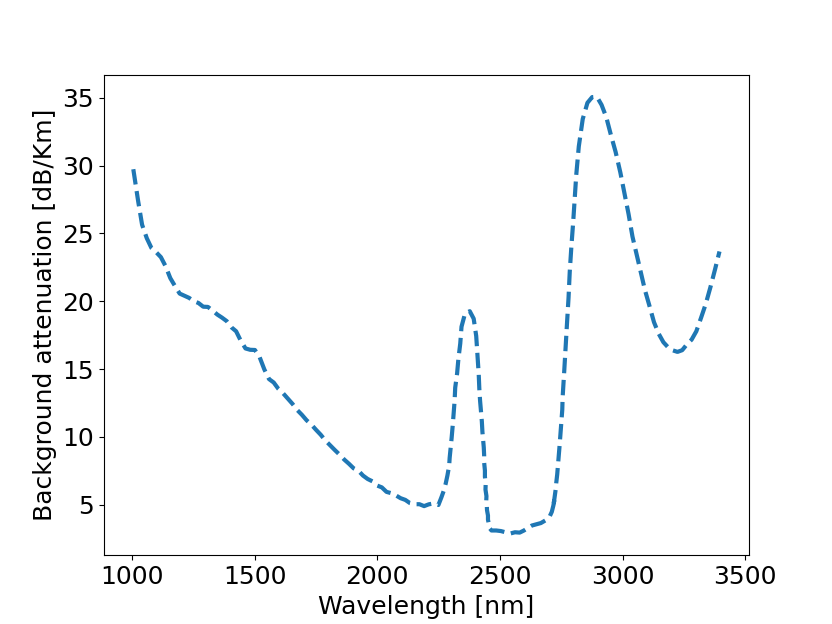}
\caption{Single mode fluoride fiber attenuation spectrum taken from LeVerreFluoré's website. A value of 35 dB/km was assumed for the pump wavelengths.}
\label{fig:atten}
\end{figure}

\begin{figure}[!t]
\centering
\includegraphics[width=3.3in]{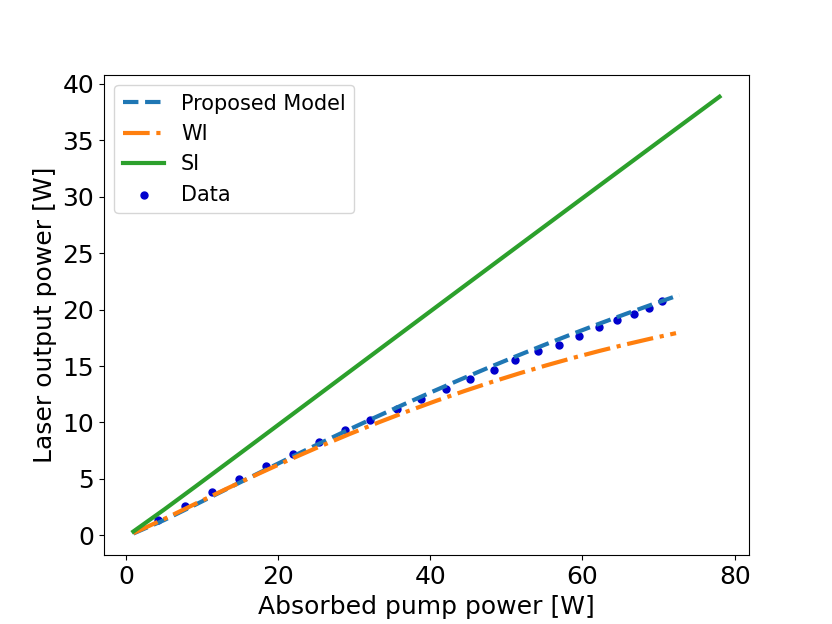}
\caption{Simulation of the laser from \cite{faucher} using the proposed model and the WI and SI parameters. The intracavity splice loss and attenuation at signal wavelength were fixed to 6.5\% and 0.1~dB/m respectively. The pump wavelength was fixed to 976~nm and the laser wavelength was 2825~nm.}
\label{fig:faucher}
\end{figure}

\begin{figure}[!t]
\centering
\includegraphics[width=3.3in]{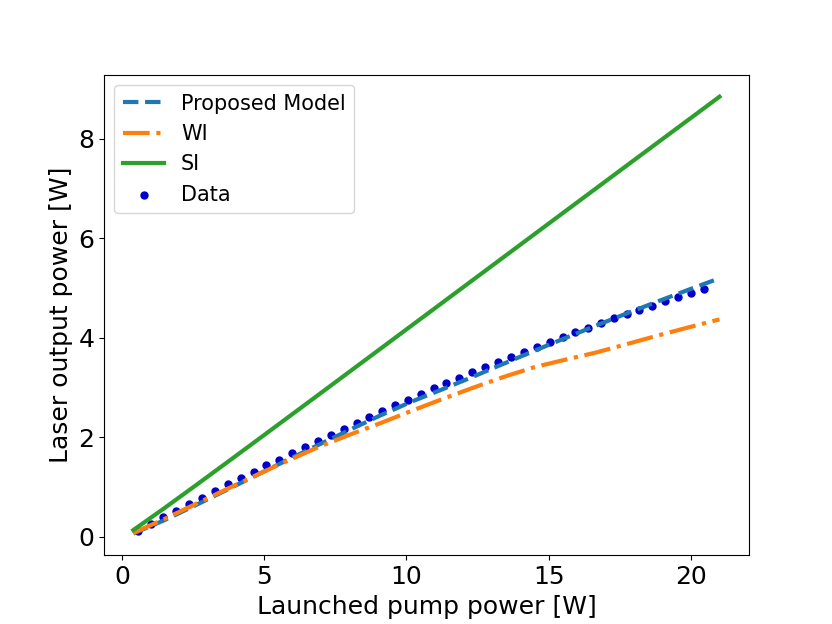}
\caption{Simulation of the laser from \cite{martin} using the proposed model and the WI and SI parameters. The attenuation at the signal wavelength was fixed to 0.17 dB/m. The pump wavelength was fixed to 976 nm, and the laser wavelength was around 2824 nm because we applied a shift to the peak Bragg wavelength as stated in ref.~\cite{martin}. Loss induced by the FBG was also considered.}
\label{fig:bernier}
\end{figure}

\begin{figure}[!t]
\centering
\includegraphics[width=3.3in]{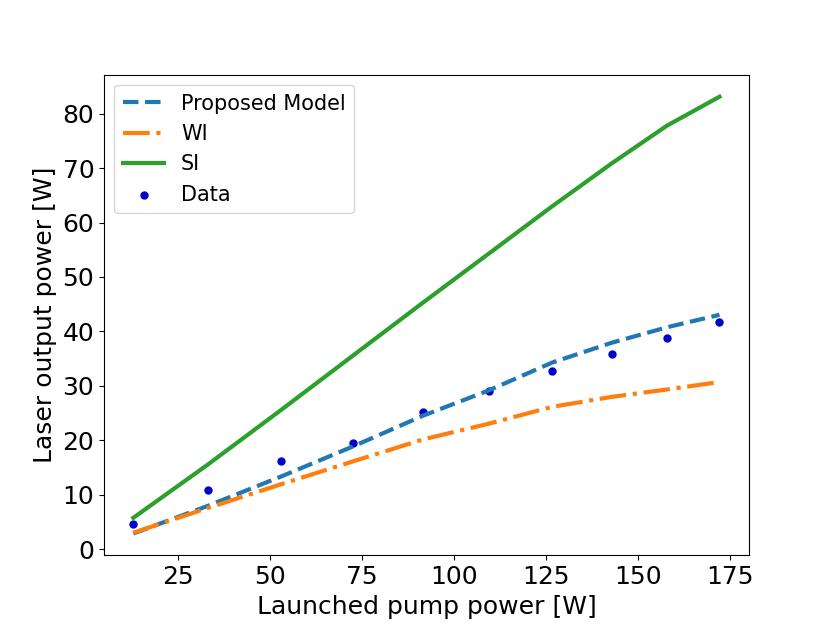}
\caption{Simulation of the laser from \cite{ozan} using the proposed model and the WI and SI parameters. The attenuation at the signal wavelength was fixed to 0.1 dB/m. The pump wavelength was changed from 969~nm to 981~nm according to the measured values \cite{ozan}, and the laser wavelength was 2825~nm.}
\label{fig:ozan}
\end{figure}

\begin{figure}[!t]
\centering
\includegraphics[width=3.3in]{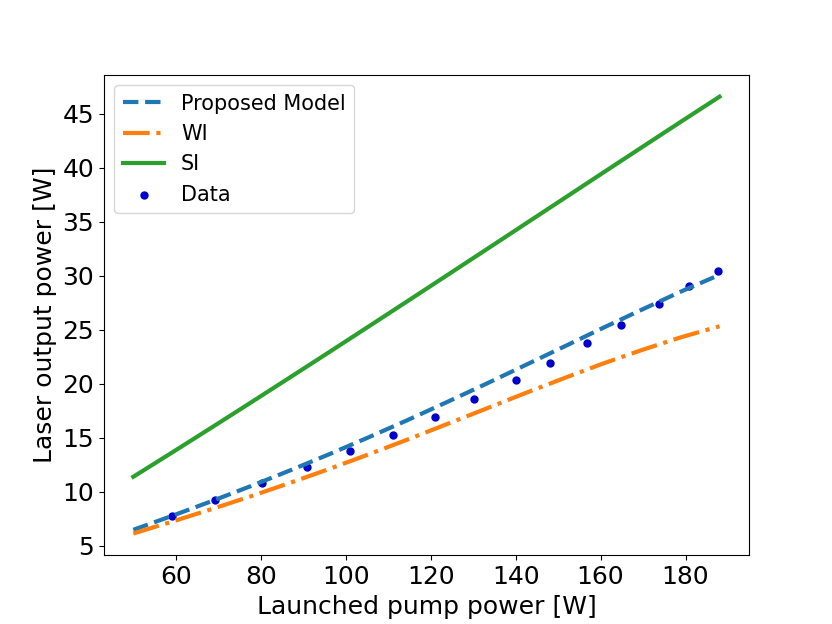}
\caption{Simulation of the laser from \cite{fortin} using the proposed model and the WI and SI parameters. The attenuation at the laser wavelength was fixed to 0.2 dB/m. The pump wavelength was changed linearly from 970~nm at 1~W to 981~nm at 200~W according to the measured values, and the signal wavelength was 2938~nm. Fusion splice loss for splices S2, S3 and S4 were fixed to 7\%, 3\% and 7\% respectively.}
\label{fig:fortin}
\end{figure}

The results obtained for the proposed  model with 4 different 7~mol.\% erbium-doped fiber lasers are shown on figures \ref{fig:faucher} to \ref{fig:fortin}. Unless otherwise specified, for the spectrum of the background loss, we used data from LeVerreFluoré \cite{lvf} for a singlemode fluoride fiber, which is shown in figure \ref{fig:atten}. For each modelled laser, this spectrum was normalized to fit the measured loss value at 2.8 $\mu$m, since they use different batches of fluoride fibers. Full spectra of the fiber Bragg gratings (FBGs) were used if available. Otherwise, Gaussian spectra were assumed.

Figure \ref{fig:faucher} shows the experimental signal power at 2825~nm vs. the absorbed pump power at 976~nm for the laser in ref.~\cite{faucher} and simulations with the three tested sets of energy transfer parameters.  With the same set of cavity parameters, the proposed model is much more accurate than the WI and SI models. In fact, we were able to reproduce the efficiency change from 35\% to 28\% with good accuracy, which was not possible with the WI and SI parameters. This efficiency change has not been reproduced that precisely before \cite{lisim}. An intracavity splice loss of 6.5\%, which is between the average value of 3.5\% and the maximum value of 9.2\% \cite{faucher}, gave the best results. 

The model was then tested on the 5~W laser from ref.~\cite{martin}. The results are shown in figure \ref{fig:bernier}. This laser has been properly simulated before, but only by taking thermal effects into account \cite{lisim}, which was not necessary in our case.

Results for the 42~W bidirectionally pumped fiber laser from ref.~\cite{ozan} are shown in figure \ref{fig:ozan}. For this laser, the pump wavelength shifted from 969~nm at low power to 981~nm at the highest power. This wavelength shift was considered in the simulation, which means different pump GSA and ESA values are used for each pump power. The simulation with the proposed model gives an efficiency that is too high up to about 125~W of launched pump power. We believe this could be due to differences between the values used for the pump GSA and ESA and their actual values for the fiber used in \cite{ozan}, which can be caused by variation in the glass composition for example \cite{csn0n2n2n6}. It could also simply be caused by experimental error in these measured cross section spectra, but their error are unknown \cite{csn0n2n2n6}. Another cause could be error in the experimental laser data or fiber properties in \cite{ozan}. Nonetheless, the predicted laser powers with the proposed model are much more accurate than the WI and SI models, especially at high pump powers as shown in figure \ref{fig:ozan}.

Results for the fourth simulated high-power 7~mol.\% laser are shown in figure \ref{fig:fortin}. The shift of the pump wavelength with power was also considered for this laser and it is believed to be the main cause of the positive efficiency shift with pump power of this laser \cite{fortin}. It was necessary to modify the cross section spectra around 2938~nm to reproduce the experimental results, which is justified since this value is in the tail of the cross section which means it has an higher uncertainty than values around 2825 nm for example \cite{csn1n2n2n1gsa}. The $\sigma_{21}$ value  taken from \cite{csn1n2n2n1gsa} was changed from $4.2 \times 10^{-26}$~m$^2$ to $3.0 \times 10^{-26}$~m$^2$ at 2938 nm, and $\sigma_{12}$ from $1.2 \times 10^{-26}$~m$^2$ to $1.5 \times 10^{-26}$~m$^2$ at the same wavelength.

\subsection{Other erbium concentrations}

 Another feature of the proposed model is that it can be used at other erbium concentrations based on the same set of energy transfer rate equations. This is evidenced in this section, where we show simulation results for four other lasers, including colasing systems, with different erbium concentrations.

\begin{figure}[!t]
\centering
\includegraphics[width=3.3in]{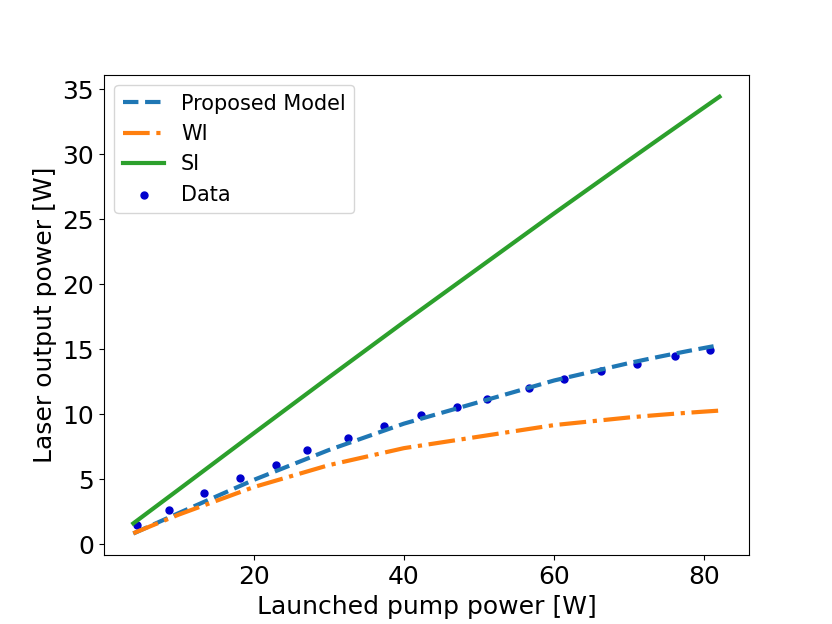}
\caption{Simulation results of an unpublished 5~mol.\% laser using the proposed model and the WI and SI parameters. The attenuation at the laser wavelength was fixed to 50 dB/km. The pump wavelength was 981~nm, and the laser wavelength was 2815~nm.}
\label{fig:Ntot5}
\end{figure}

To show that the proposed model can indeed be used with different erbium concentrations, it was first applied to an unpublished result obtained for a  2815~nm fiber laser pumped at 981~nm. The erbium-doped fluoride fiber had an erbium concentration of 5~mol.\%, a core radius of 5~$\mu$m, a cladding radius of 125~$\mu$m and a NA of 0.188. The active fiber length was 14.5~m, the low reflectivity ouput coupler FBG (LRFBG) had a  reflectance of 7.1\% at 2815~nm and the high reflective input coupler FBG (HRFBG) had a reflectance of 96.5\% at the same wavelength. The output end of the  laser was terminated with an AlF$_3$ endcap. The simulation results are showing very good agreement with experimental data as can be seen in figure \ref{fig:Ntot5}.
In obtaining the results of figure \ref{fig:Ntot5}, equations \eqref{eq:fulletu1} to \eqref{eq:fullcr} were used without any modification, i.e. based on the very same set of parameters.

\begin{figure}[!t]
\centering
\includegraphics[width=3.3in]{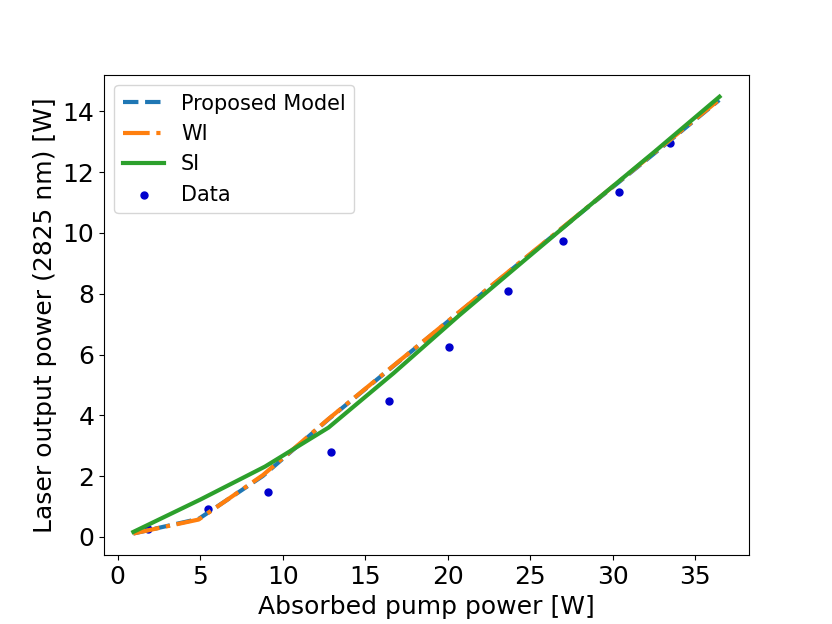}
\caption{Simulation results of the 2825~nm signal of the laser from \cite{ozancolasing} using the proposed model and the WI and SI parameters. The attenuation was fixed to 30 dB/km at 2825 nm. The pump wavelength was fixed to 976~nm. Loss values for all the components were set to the values stated in ref. \cite{ozancolasing} without any modification.}
\label{fig:2825}
\end{figure}

Following the work by Guo et al. \cite{guo}, we attempted to reproduce the experimental results from the laser in ref.~\cite{ozancolasing} by taking the ESA at 1.6~$\mu$m from level 1 to 3 into account. Note that this laser used a significantly lower erbium concentration of 1~mol.\%. Our simulation results along with experimental data are shown in figure \ref{fig:2825}. 
Results for the 2825~nm signal are for this case in less good agreement with the experimental data, as the laser efficiency is a bit too low. Also, at such low erbium concentrations, the energy transfer model has a much lower impact on predicted laser power than for the 5 mol.\%, 6 mol.\% and 7 mol.\% lasers, which can be seen on figure \ref{fig:2825}. The WI model and the proposed model both give very similar results.

\begin{figure}[!t]
\centering
\includegraphics[width=3.3in]{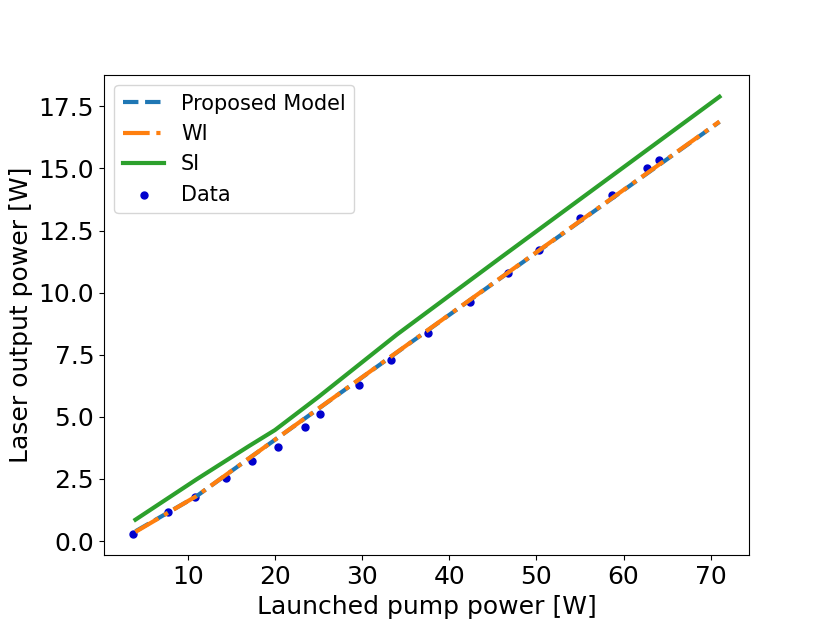}
\caption{Simulation results of the 2.8 $\mu$m signal of the  laser from \cite{licolasing} using the proposed model and the WI and SI parameters. The attenuation spectrum for FiberLabs monomode fluoride fibers was used, which is 17 dB/km at 976~nm, 30~dB/km at 1613~nm and 43~dB/km at 2887~nm. The pump wavelength was fixed to 976~nm. Loss values for all the components were set to the values stated in the article without any modification. The filter loss was set to 20\%.}
\label{fig:licolasing28}
\end{figure}

A colasing system using a 1.5~mol.\% fiber from ref.~\cite{licolasing} was also simulated. The main difference with the other studied lasers in this work is the use of broadband mirrors instead of the narrowband FBGs. For such lasers, a redshift of the laser wavelength with increasing pump power is usually observed \cite{gorjan, lisim}. Our simulation results along with experimental data are shown in figure \ref{fig:licolasing28}.
We observed good agreement between our predicted laser powers and the experimental results.
The proposed model uses a spectrum of evenly distributed wavelength channels, so it was possible to observe a redshift in the output spectrum of the simulation.  We observed wavelengths of 2871~nm at 4~W of pump power and 2887~nm  at higher pump powers. The predicted redshift vs. pump power is much stronger in our simulation than in reality, which could be due to error in the measured cross section spectra of the 2.8~$\mu$m laser transition, which in turn affects the peak gain value at each pump power.

\begin{figure}[!t]
\centering
\includegraphics[width=3.3in]{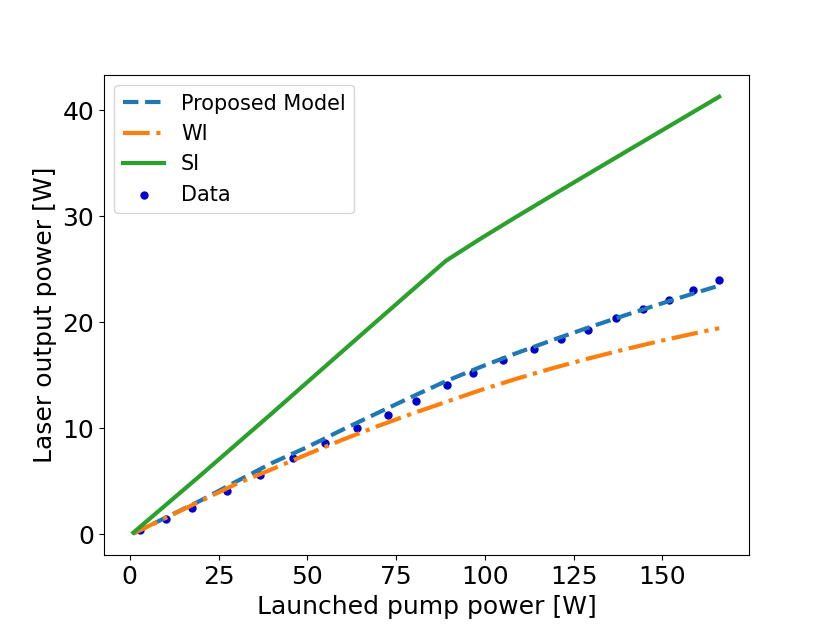}
\caption{Simulation of the laser from \cite{tokita} using the proposed model and the WI and SI parameters. The attenuation was fixed to 0.55 dB/m at 2887 nm.  The pump wavelength is 975~nm, and the laser wavelength ranges from 2802~nm to 2887~nm. Intracavity signal re-injection loss of 40\% and pump injection loss of around 40\% on both sides of the cavity were used.}
\label{fig:tokita}
\end{figure}

Finally, as shown in figure \ref{fig:tokita}, we also reproduced results from the laser in ref.~\cite{tokita} which correspond to a 2.8~$\mu$m laser also involving broadband mirrors and a 6 mol.\% erbium doped fiber.
We needed to consider very high intracavity losses to properly simulate  this laser, and no information on these losses can be found in ref. \cite{tokita}.  The simulated lasing wavelength was 2802 nm at 1 W of pump power, 2824~nm at 20, 30 and 40~W, 2871~nm at 50 and 60~W and 2887~nm at higher pump powers. It cannot be compared to ref.~\cite{tokita} because no information on the precise lasing wavelength is given.

\subsection{Transfer rate analysis}

\begin{figure}[!t]
\centering
\includegraphics[width=3.3in]{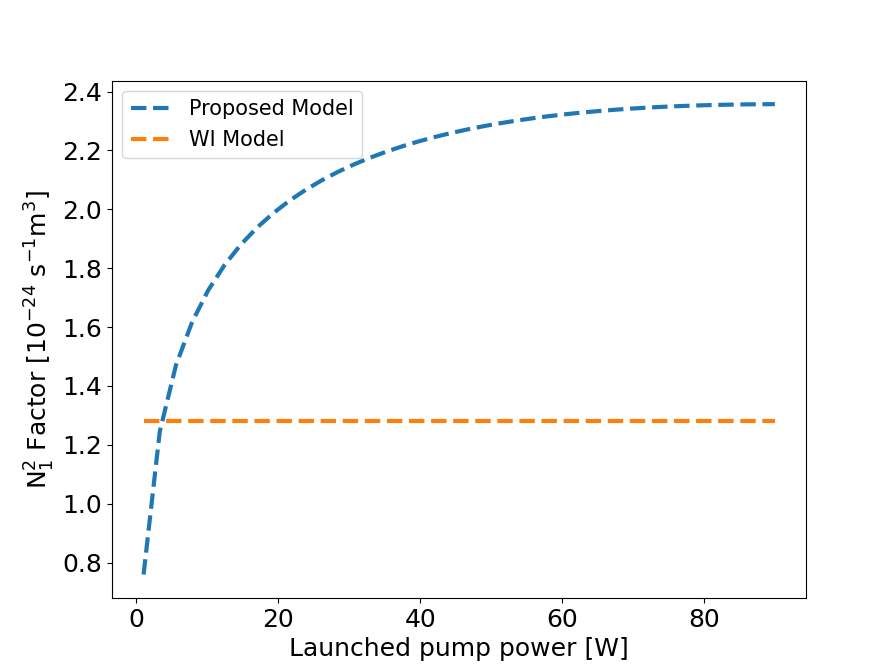}
\caption{Factor in front of $N_1^2$ for ETU1 in the proposed model and the WI model vs launched pump power at the fiber entry (pump side) calculated using the simulation of the laser shown in figure \ref{fig:faucher}.  Atomic populations of levels 0 and 1 were taken from figure \ref{fig:atpop}.}
\label{fig:hopping}
\end{figure}

\begin{figure}[!t]
\centering
\includegraphics[width=3.3in]{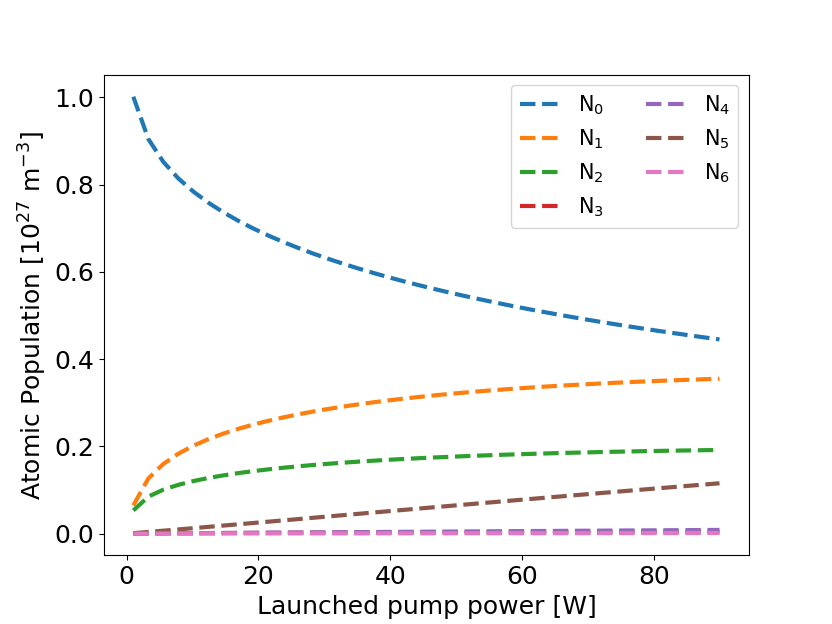}
\caption{Atomic population vs. pump power at the fiber entry (pump side) for the simulation results with the proposed model shown in figure \ref{fig:faucher}. }
\label{fig:atpop}
\end{figure}

Figure \ref{fig:faucher} shows that the proposed model predicts signal powers that are similar to those predicted with the WI model at low pump powers, while it is much more accurate at higher pump powers. Accordingly, it is interesting to follow the ETU1 transfer rate, which has a direct impact on lasing efficiency, of the proposed model as compared to that of the WI model.  This comparison is illustrated in figure \ref{fig:hopping} where we show how the factor in front of $N_1^2$ changes with the input pump power, whereas figure \ref{fig:atpop} shows the corresponding atomic population evolutions. Equation \eqref{eq:etu1grant} is used for the WI model and equation \eqref{eq:fulletu1} is used for the proposed model.
The factor in front of $N_1^2$  is close to the WI value at low pump power, which is expected since the resulting signal powers in figure \ref{fig:faucher} for both models are close at such low powers. However, the factor quickly rises with pump power to reach a near constant value of about 2 times the WI parameter, which explains why the proposed model leads to higher signal powers at these pump power values. This behavior of the transfer rate is attributed to the shape of the $N_1$ curve shown in figure \ref{fig:atpop}. 

The proposed model can also easily be used to add energy transfer processes to the atomic population rate equations, as long as the relevant cross sections to compute the overlap integrals in equation \eqref{eq:microp} are known. 
For example, the cross relaxation processes CR3 and CR6, which are the opposite of ETU1 and ETU2 respectively, were tested. These were tested before in ref.~\cite{gorjan}, but the transfer rates were unknown. In our case, they could easily be computed using the overlap integrals between $\sigma_{31}$ and $\sigma_{01}$ for CR3 and between $\sigma_{62}$ and $\sigma_{02}$ for CR6. The McCumber relation was used to get the cross sections that were unknown. The overlap value is  $1 \times 10^{-59}$~m$^5$ for CR3 and   $1 \times 10^{-57}$~m$^5$ for CR6. However, our simulations showed that their effect on laser efficiency is negligible since the atomic populations of levels 3 and 6 are very low which confirms the results from Gorjan et al. \cite{gorjan}.

\section{Conclusion}

In summary, we have shown that the WI and SI models which were proposed to model high-power erbium-doped fluoride fiber lasers do not reproduce well the highly doped fiber laser powers and efficiencies. We proposed a cubic dependence of the energy transfer rate and found that combining both Dexter's model and the hopping model of migration assisted energy transfer gave the best results. Eight high-power erbium-doped fluoride fiber lasers operating around 2.8~$\mu$m  were simulated to prove that the proposed model gives better agreement with experimental data for multiple erbium concentrations. The efficiency drop which is generally observed at high pump powers was reproduced with good accuracy, which means the developed model can be used for accurate power scaling predictions. 
However, there is still place for improvement. For example, we believe it should be possible to consider  more general energy transfer equations which are not the combination of the extreme cases of energy transfer without migration and purely migration assisted energy transfer. In future works, the proposed model will be applied to different mid-infrared fiber laser systems, like the 3.5 microns erbium system for example \cite{maxime}. We believe the proposed model will help designing the next generation of erbium-doped fluoride fiber and the cavity parameters needed to produce average power in excess of 100W around 2.8 microns.

\section*{Acknowledgments}
The authors would like to thank Fonds de recherche du Québec – Nature et technologies (144 616, CO256655), Canada Foundation for Innovation (5180) and Natural Sciences and Engineering Research Council of Canada (IRCPJ469414-13, RGPIN-2016- 05877) for financing this work, as well as Vincent Fortin for helpful discussions and technical assistance.

\appendix

The starting point of the first terms in equations \eqref{eq:fulletu1} to \eqref{eq:fullcr} is equation \eqref{eq:transfertrate} which gives the transfer rate between two ions separated by a distance $d$ for a dipole-dipole interaction. 
Dai et al. \cite{dai} have proposed to relate the total erbium concentration in the fiber to the inverse of the volume of a sphere of radius equal to the interaction distance $d$. Following this approach, we rather propose to relate $d$  to the geometrical mean of ${N_i, N_j}$ via the following equation:
\begin{align}
    d = \qty(\frac{4 \pi}{3})^{-1/3} (\sqrt{N_i N_j})^{-1/3} \equiv a (N_i N_j)^{-1/6},
    \label{eq:linkdistance}
\end{align}
where $N_i$ and $N_j$ are the concentration of the levels $i$ and $j$ respectively involved in a given energy transfer process. We believe using the excited level concentration instead of the total concentration is more appropriate since it is the excited ions that actually transfer their energy. Substituting \eqref{eq:linkdistance}  into \eqref{eq:transfertrate} and assuming a cubic concentration for the  transfer rate in units of m$^{-3}$s$^{-1}$ gives:
\begin{align}
     f_{1} = a^{-6} C_{etu, cr} (N_i N_j)^{3/2}.
     \label{eq:directratesunits}
\end{align}
 Equation \eqref{eq:directratesunits} with the appropriate populations $N_i$ and $N_j$ corresponds to the first term in equations \eqref{eq:fulletu1} to \eqref{eq:fullcr}.

The starting point of the second terms of equations \eqref{eq:fulletu1} to \eqref{eq:fullcr} is the following equation \cite{bartolo}:
\begin{align}
    W_{Hopping} = \sqrt{\frac{\pi}{6}} a^{-6} \sqrt{C_{DA} C_{DD}} N_A N_D.
    \label{eq:hopping}
\end{align}
Equation \eqref{eq:hopping} represents the transfer rate of a migration-assisted transfer process in which the migration is characterized by the hopping mechanism. It was adapted from equation 287 from ref. \cite{bartolo} to consider its dependence in $a$. $C_{DA}$ and $C_{DD}$ are the Förster microparameters for ETU or CR and migration respectively. In our notation, this corresponds to $C_{DA} \rightarrow C_{etu, cr}$ and $C_{DD} \rightarrow C_{m}$. $N_D$ is the donor concentration and is linked to the ions undergoing the migration process. $N_A$ the acceptor concentration and is linked  to the ions undergoing the CR or ETU process which is assisted by the migration process. For a transfer process between two ions starting at the same excited level $i$, we thus propose the following transfer rate in units of m$^{-3}$s$^{-1}$:
\begin{align}
    f_2 = \sqrt{\frac{\pi}{6}} a^{-6} \sqrt{C_{etu, cr} C_{m}} N_0^{1/4} N_i^{11/4},
    \label{eq:hoppingmodunits}
\end{align}
where the concentration dependence was found empirically. The $N_0$ dependence is justified by the $N_D$ dependence of equation \eqref{eq:hopping} and is necessary since no migration can occur if $N_0 \rightarrow 0$.
For a process where one of the ions starts at the fundamental level, we propose the following transfer rate:
\begin{align}
    f_2 = \sqrt{\frac{\pi}{6}} a^{-6} \sqrt{C_{etu, cr} C_{m}} (N_0 N_i)^{3/2},
    \label{eq:hoppingmodunitsNiNj}
\end{align}
which is approximately the same as applying $N_i \rightarrow \sqrt{N_0 N_i}$ to equation \eqref{eq:hoppingmodunits}.  Equation \eqref{eq:hoppingmodunits} and \eqref{eq:hoppingmodunitsNiNj} with the appropriate populations $N_i$ and $N_j$ corresponds to the second term in equations \eqref{eq:fulletu1} to \eqref{eq:fullcr}.

Two more points have to be clarified. Firstly, a cubic energy transfer rate comes from the assumption that the excited ions are uniformly distributed in the system \cite{wangsimkin}. On the other hand, a quadratic dependence is obtained by assuming a random distribution of ions and an infinitely fast energy migration process \cite{grant}.

Finally, it appeared that adding both terms $f_1$ and $f_2$ was necessary in order to obtain transfer rate values leading to accurate and consistent simulation results in various situations. The first term is the transfer without the assistance of migration and the second term is the transfer purely assisted by migration \cite{jacksonchap6}. The need to add both terms might indicate that the corresponding physical processes  both occur in the erbium-doped laser system.

\begin{IEEEbiography}[{\includegraphics[width=1in,height=1.25in,clip,keepaspectratio]{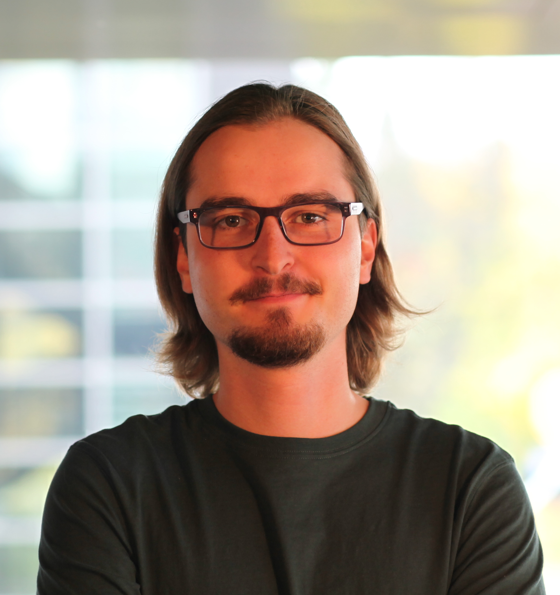}}]{William Bisson}
received a B. Sc.  degree in physics from Laval University, Quebec City, Qc, Canada in 2022.   He is currently pursuing his M. Sc. in physics at the Center of Optics, Photonics and lasers. His research interests include modeling of mid-infrared erbium-doped fluoride fiber lasers and neodymium silica fiber lasers.
\end{IEEEbiography}
\begin{IEEEbiography}[{\includegraphics[width=1in,height=1.25in,clip,keepaspectratio]{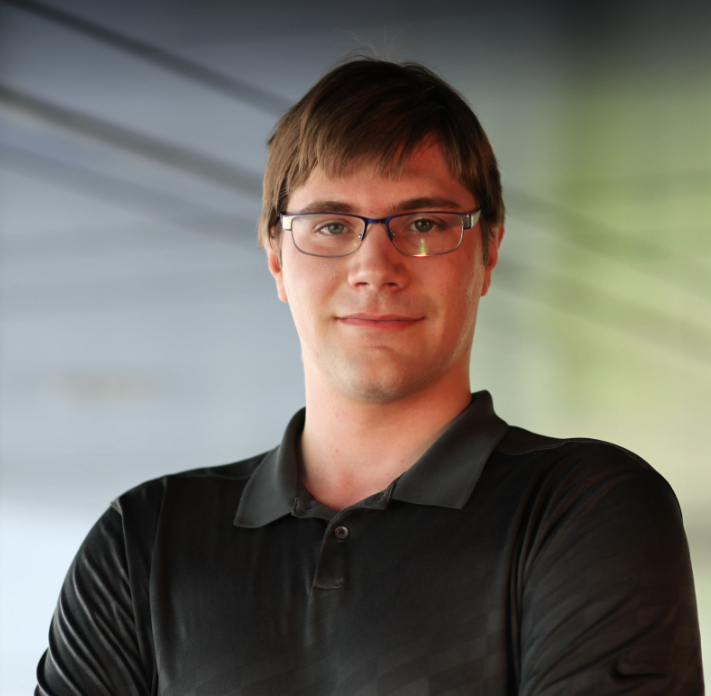}}]{Alexandre Michaud}
received a B. Eng.  degree in physical engineering from Laval University, Quebec City, Qc, Canada in 2022.  He is currently pursuing his M. Sc. in physics at the Center of Optics, Photonics and lasers. His research interests include the development of Raman fiber lasers in the mid-infrared and neodymium silica fiber lasers.
\end{IEEEbiography}
\begin{IEEEbiography}[{\includegraphics[width=1in,height=1.25in,clip,keepaspectratio]{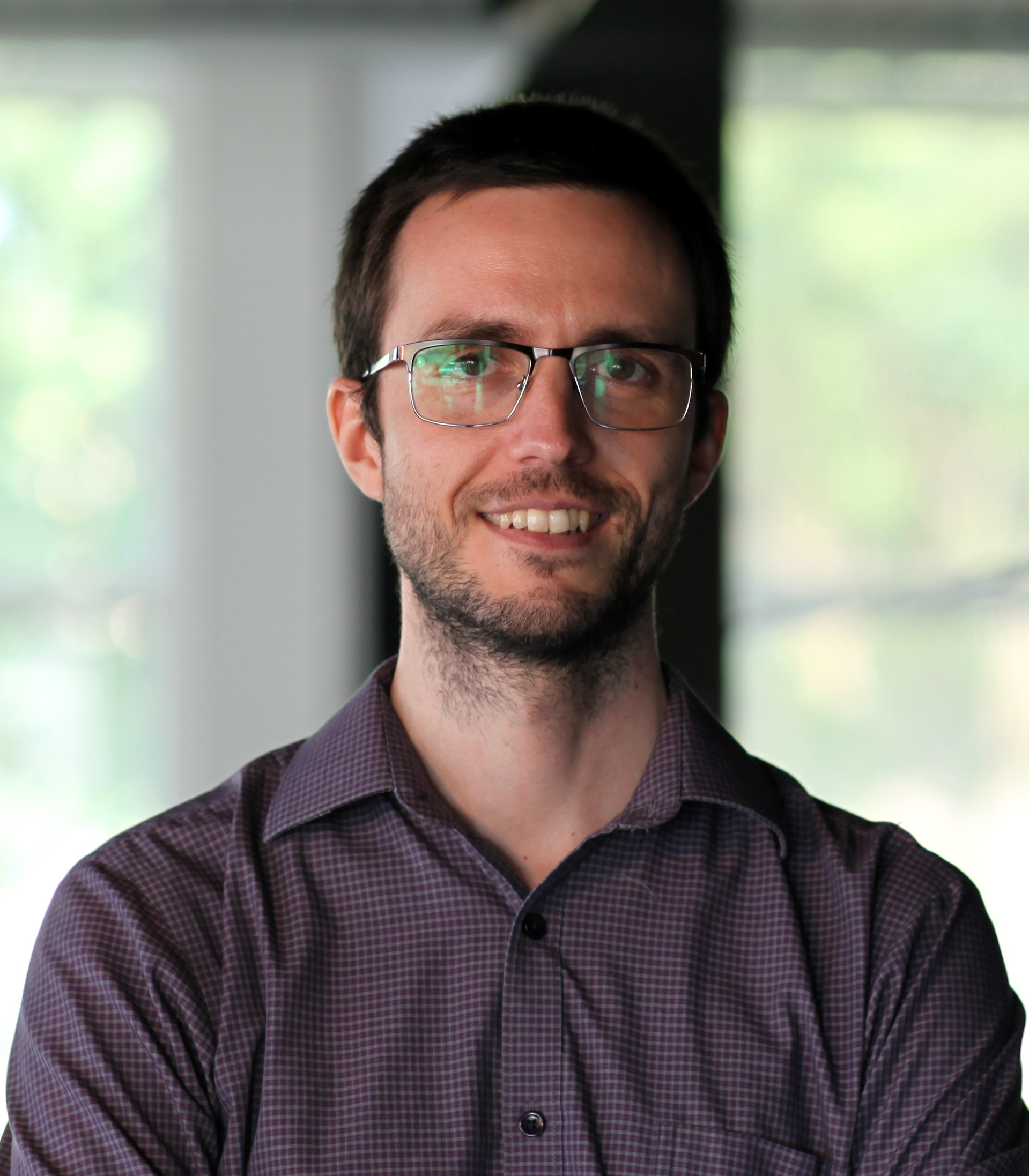}}]{Pascal Paradis}
received a B. Eng. degree in engineering physics, a masters degree and a Ph. D. degree in Physics from Laval University, Quebec City, Qc, Canada in 2017, 2019 and 2023. He is currently a postdoctoral fellow at the Center for Optics, Photonics and Lasers, Laval University. His research interests include pulsed mid-infrared fiber lasers for spectroscopy, remote sensing, material processing and biomedical applications.
\end{IEEEbiography}
\begin{IEEEbiography}
[{\includegraphics[width=1in,height=1.25in,clip,keepaspectratio]{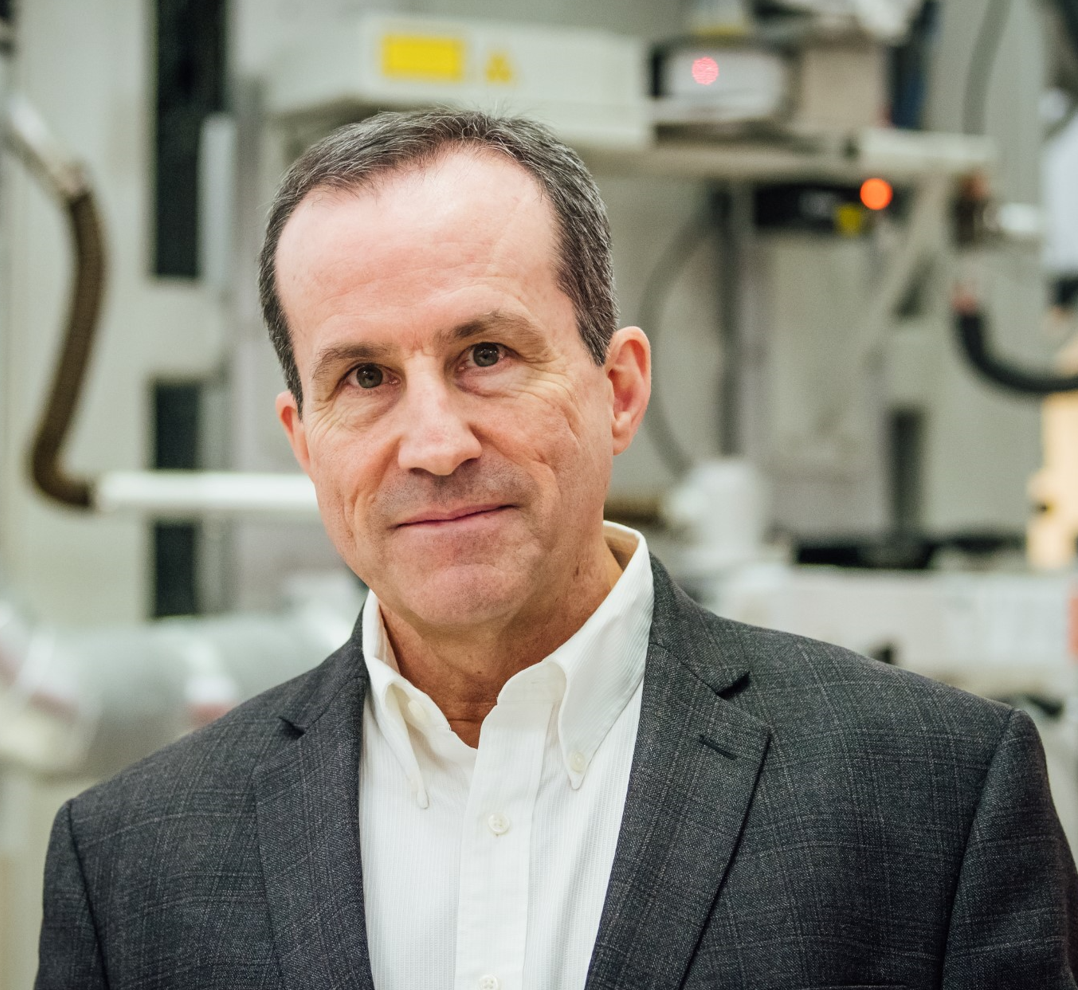}}]{Réal Vallée}
received his Ph.D. from Laval University in 1986. He was a postdoctoral fellow at the Laboratory for Laser Energetics, at the University of Rochester before returning to Laval University as a professor in the Department of Physics. In 2000 he was appointed Director of the Center for Optics Photonics and Lasers (COPL), the province of Quebec network center of excellence in photonics, and held this position until 2020. His research interests are in fiberoptics components and their applications, non-linear propagation of ultrafast pulses in fiber, inscription of waveguides with femtosecond pulses and the study of infrared glasses for integrated optics. He currently holds an NSERC-funded Industrial Research Chair in Femtosecond Photo-Inscribed Photonic Components and Devices. Professor Vallée has supervised over 100 graduate students, has authored over 200 peer-reviewed papers in high impact journals and holds 16 patents. He successfully spearheaded a grant application to the Canada Foundation for Innovation for the construction of a building entirely dedicated to optics and photonics university research and training. He was promoted Fellow of the Optical Society of America in 2011 and is the recipient of the Brockhouse and Urgel-Archambault prizes.
\end{IEEEbiography}
\begin{IEEEbiography}[{\includegraphics[width=1in,height=1.25in,clip,keepaspectratio]{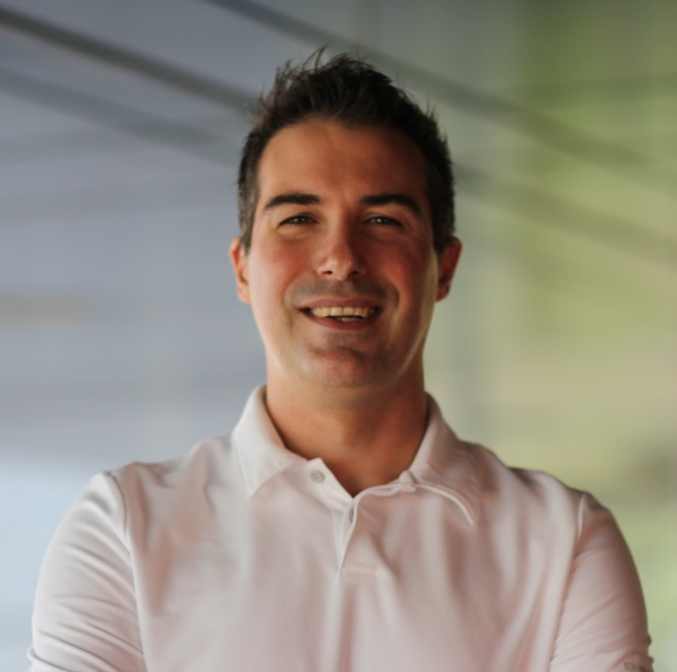}}]{Martin Bernier}
received the Ph.D. degree in 2010 in physics from Laval University, Quebec City, QC, Canada, where he is currently Full Professor with the Department of Physics, Engineering Physics and Optics. His research projects conducted with the Center for Optics, Photonics and Lasers, Laval University, involve writing of fiber Bragg gratings using femtosecond pulses and their application to the development of innovative fiber-based components, particularly in the field of fiber lasers and sensors.
\end{IEEEbiography}

\newpage

\end{document}